\documentclass[12pt]{article}
\usepackage{pic04}
\usepackage{hyperref}
\usepackage{url}
\usepackage{graphicx}

\def\EPJ                     {\mbox{Eur. Phys. J.}}
\def\JHEP                    {\mbox{JHEP}}

\def\PhysLett                {\mbox{Phys. Lett.}}

\def\PhysRev                 {\mbox{Phys. Rev.}}
\def\PRL                     {\mbox{Phys. Rev. Lett.}}

\newcommand{\Journal}[4]     {{#1} {\bf {#2}}, {#3} ({#4})}

\newcommand{\WWWAddr}[1]     {{\tt {#1}}}
\newcommand{\hepref}[2]      {{\tt hep-{#1}/{#2}}}

\newlength{\onemp}
\newlength{\twomp}
\newlength{\threemp}
\newcommand{\onefig}[1]{%
  \begin{center}
  \includegraphics[width=\onemp]{#1}
  \end{center}
}
\newcommand{\twofigs}[2]{%
  \begin{center}
  \begin{minipage}[b]{\twomp}
    \centering
    \includegraphics[width=\textwidth]{#1}
  \end{minipage}
  \begin{minipage}[b]{\twomp}
    \centering
    \includegraphics[width=\textwidth]{#2}
  \end{minipage}
  \end{center}
}
\newcommand{\threefigs}[3]{%
  \begin{center}
  \begin{minipage}[c]{\threemp}
    \centering
    \includegraphics[width=\textwidth]{#1}
  \end{minipage}
  \begin{minipage}[c]{\threemp}
    \centering
    \includegraphics[width=\textwidth]{#2}
  \end{minipage}
  \begin{minipage}[c]{\threemp}
    \centering
    \includegraphics[width=\textwidth]{#3}
  \end{minipage}
  \end{center}
}

\newcommand{\etal}      {\mbox{\it et al.}}
\newcommand {\Dzero}    {\mbox{D{\O}}}

\newcommand {\unitexp}[2] {\mbox{{#1}$^{\mathrm{#2}}$}}

\newcommand {\ra}         {\mbox{$\rightarrow$}}

\newcommand {\pT}       {\mbox{$p_T$}}
\newcommand {\ET}       {\mbox{$p_T$}}
\newcommand {\Qsq}      {\mbox{$Q^2$}}
\newcommand {\Ptrel}    {\mbox{$P_t^{\mathrm{rel}}$}}
\newcommand {\alphas}   {\mbox{$\alpha_s$}}

\newcommand {\qrk}[1]   {\mbox{$#1$}}
\newcommand {\aqrk}[1]  {\mbox{$\bar{#1}$}}

\newcommand {\Hadron}[2]  {\mbox{${#1}_{#2}$}}
\newcommand {\Hadronup}[2]{\mbox{${#1}^{#2}$}}
\newcommand {\Xpart}      {\mbox{$X$(3872)}}
\newcommand {\pentaq}     {\mbox{$\Theta^+$}}
\newcommand {\cpentaq}    {\mbox{$\Theta_c^0$}}

\newcommand {\qqb}[1]   {\mbox{\qrk{#1}\aqrk{#1}}}
\newcommand {\ep}       {\mbox{$ep$}}
\newcommand {\epem}     {\mbox{$e^+e^-$}}
\newcommand {\mupmum}   {\mbox{$\mu^+\mu^-$}}

\newcommand {\pim}      {\mbox{$\pi^-$}}
\newcommand {\pip}      {\mbox{$\pi^+$}}

\newcommand {\Jpsi}     {\mbox{$J/\psi$}}
\newcommand {\psipr}    {\mbox{$\psi$(2S)}}
\newcommand {\Ups}      {\mbox{$\Upsilon$}}
\newcommand {\xicc}     {\mbox{$\Xi_{cc}^+$(3520)}}
\newcommand {\DsJ}      {\mbox{$D_{sJ}^+$(2632)}}

\begin{document}

\title{\bf HEAVY FLAVOR PRODUCTION IN HADRON COLLISIONS
            (WITH A FEW LEPTONS AND PHOTONS THROWN IN)}
\author{
        Harold G. Evans        \\
       {\em Columbia University, New York, NY, USA}}
\maketitle

%
\vspace{4.5cm}

\baselineskip=14.5pt
\begin{abstract}
Substantial advances in our understanding of several aspects of QCD
have been achieved in the recent past using heavy quarks as a tool.
However, many open questions still remain.
These successes and puzzles are highlighted by
the latest measurements of heavy quark production at the
Tevatron, HERA and fixed target experiments, which will be reviewed
here.
Results in both open heavy flavor and heavy quarkonium production as
well as evidence for new particles containing heavy quarks will be
presented.
The impact of these measurements on gaps in our understanding of QCD
and how we hope to close these gaps in the future will be outlined.
\end{abstract}
\newpage

\baselineskip=17pt

\section{Why Study Heavy Quark Production?}
Quantum Chromodynamics is universally acknowledged to be {\em the}
theory of the strong force. However, its study continues to be a
compelling area of research because of the difficulty of performing
calculations in regions where the theory becomes
non-perturbative. This means that, although we understand the
structure of the theory, we still cannot make accurate predictions for
a wide range of important observables. Intellectually, this is
frustrating (or an opportunity for the more optimistic). But it also
has a more practical consequence. Our understanding of QCD processes
is intimately entwined with our understanding of other aspects of the
Standard Model because QCD is a part of all SM predictions, from
estimates of backgrounds to corrections to electro-weak observables.

It turns out that the production of heavy quarks by the strong force
is an excellent area to study some of the technical details of QCD
that are so important in our tests of the Standard Model.
To understand this, consider
the production of a heavy quark-antiquark pair
in the collision of two particles.
Broadly speaking, this process consists of three components, which are
all connected in real collisions: the structure of the incoming
particles, the hard interaction producing the \qrk{Q}\aqrk{Q} pair and
the subsequent parton shower and fragmentation of the final state
partons to produce observable hadrons.

It is in the second of these
entwined processes, the hard scattering, where heavy quarks make their
contribution to QCD. Particle structure and hadronization are clearly
governed by non-perturbative physics. However, they are also largely
universal functions, appearing in a variety of processes. The
hard-scatter is process dependent. But since
the masses of heavy quarks are much larger than the QCD scale,
this hard-scatter should be calculable
using perturbative QCD. Heavy quark production measurements can
therefore be used to probe our ability to do perturbative calculations
or can be used as a tool to understand parton densities and
fragmentation.

Before embarking on a discussion of specific heavy quark production
results, we should be clear as to exactly what a heavy quark is. In
this paper, heavy quarks are taken to be \qrk{b}- and
\qrk{c}-quarks. The obese \qrk{t}-quark is discussed in a separate
contribution to these proceedings \cite{hocker}.
Using this definition, heavy quarks are produced at a variety of
facilities. A comparison of those for which results are presented
is given in Table \ref{table:facilities}.

\begin{table}[htb]
\begin{center}
  \caption{ \it Comparison of experimental facilities with results
  presented here.} 
  \vskip 0.1 in
  \begin{tabular}{|c|cc|c|c|c|} 
  \hline
    {\bf Exp. or} & \multicolumn{2}{c|}{\bf Colliding}
      & $\mathbf{\sqrt{s}}$ {\bf / nucl.} 
      & {\bf Runs} & {\bf Recorded} \\
    {\bf Facility} & \multicolumn{2}{c|}{\bf Particles}
      & {\bf [GeV]}
      & & {\bf Data} \\
  \hline
  \hline
    FOCUS/E831
      & $\gamma$ & BeO     & 18 & 96--97 & \\
    SELEX/E781
      & $\Sigma^-,\pi^-$ & C,Cu & 33 & 96--97 & 15B int's \\
    NuSea/E866
      & $p$   & Cu      & 38 & 96--97 & 9M \Jpsi \\
    Hera-B
      & $p$   & C,Al,Ti,W & 41 & 00,02--03 & 308K \Jpsi \\
  \hline
    HERA Run I
      & $e^{\pm}$ & $p$ & 300,318 & 93--00 & 130 \unitexp{pb}{-1} \\
    HERA Run II
      & $e^{\pm}$ & $p$ & 318 & 03--04 & $\sim$70 \unitexp{pb}{-1} \\
  \hline
    Tevatron Run I
      & $p$   & $\bar{p}$ & 1800 & 92--96 & 125 \unitexp{pb}{-1} \\
    Tevatron Run II
      & $p$   & $\bar{p}$ & 1960 & 02--04 & $\sim$400 \unitexp{pb}{-1} \\
  \hline
    LEP (I and II)
      & $e^+$ & $e^-$ & 90--210 & 89--00 & 3.6M \qqb{b} \\
  \hline
  \end{tabular}
  \label{table:facilities}
\end{center}
\end{table}

\section{\qrk{b}- and \qrk{c}-Quark Production}
\subsection{History Lessons}
The production of open \qrk{b}- and \qrk{c}-quarks has been one of the
most troubling problems in QCD for more than a decade. For a recent
review of this problem see \cite{cacciari}.
Particularly in
the \qrk{b}-quark sector, calculations, which are done at
next-to-leading order (NLO) in \alphas , 
were expected to provide a quite good description of the data. 
A quick look at the data taken prior to 2000 \cite{cacciari},
however,
indicates that while the shape of the \qrk{b}-quark production
cross-section is reasonably well modeled by NLO theory for \qqb{p}\ra
\qqb{b}, e\qrk{p}\ra \qqb{b} and $\gamma\gamma$\ra \qqb{b} data, the
predictions underestimate the magnitude of the 
cross-sections by
factors approaching three.
Surprisingly, data and predictions for \qrk{c}-quark production showed
much better agreement, although with larger uncertainties.

Over the past few years, the picture of 
\qrk{b}-quark production at the Tevatron,
where the discrepancy was originally uncovered,
has become much clearer.
One important aspect of this understanding
was the realization that experimentalists should report what they observe:
\Hadron{B}{}-hadron production cross-sections, rather than
cross-sections corrected to the \qrk{b}-quark level. When the \Dzero\
collaboration published a measurement of the \ET\ distribution of jets
containing \qrk{b}-quarks \cite{d0bjet}, significantly better
agreement with NLO predictions was found.
Another piece of the puzzle was the correct incorporation of
next-to-leading-log resummation of log(\pT /$m$) terms with the NLO
hard scatter calculation including massive quarks (FONLL) \cite{fonll}.
Finally, the heavy quark fragmentation function was revisited by
several groups \cite{bkk,caccnas}
yielding a calculation in the FONLL framework consistent with the hard
scattering calculation and a reevaluation of parameters of the
fragmentation function.

\subsection{Open Beauty and Charm Production at the Tevatron}
\begin{figure}[htb]
  \begin{center}
  \begin{minipage}[b]{0.44\textwidth}
    \centering
    \includegraphics[width=\textwidth]{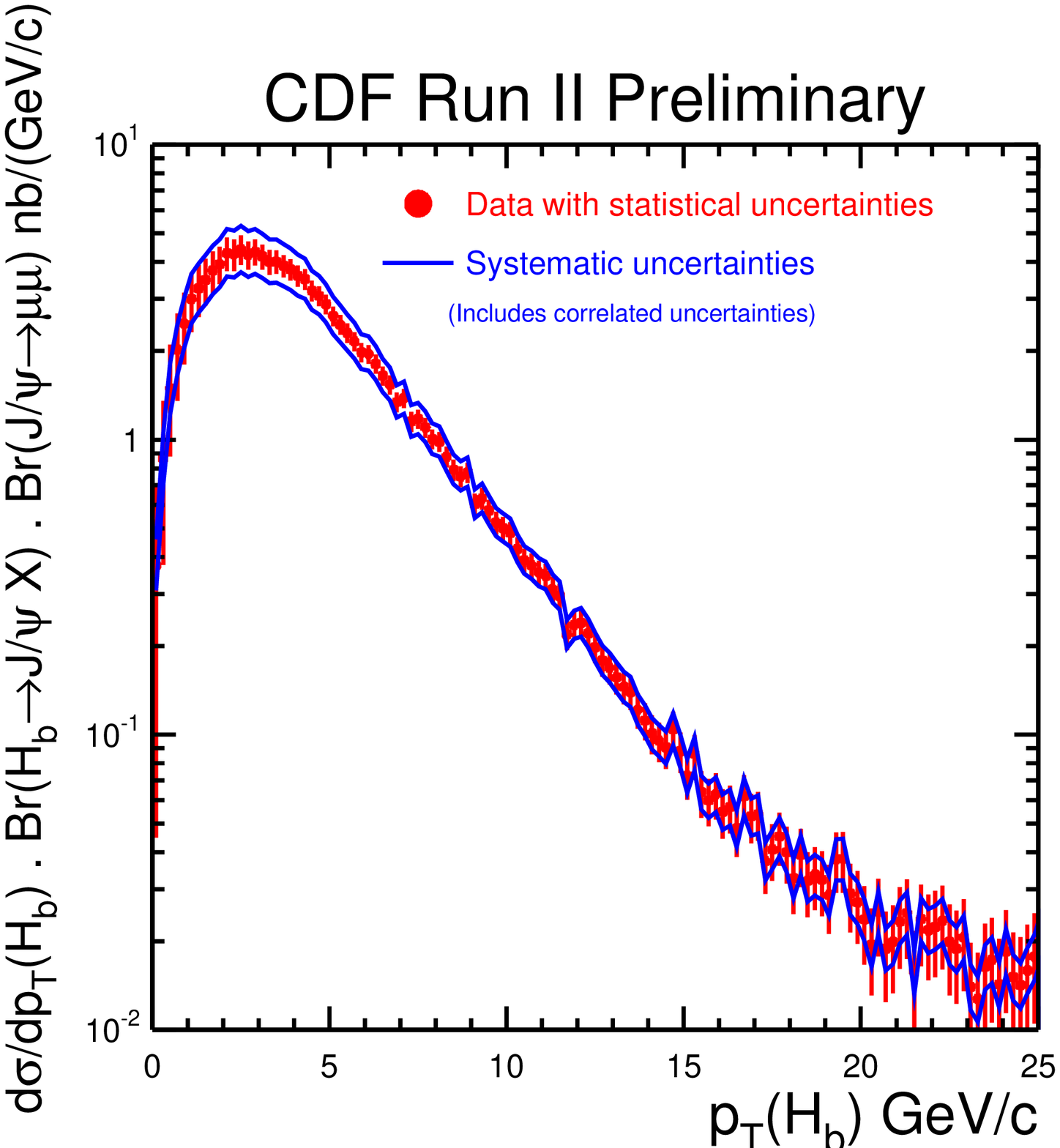}
  \end{minipage}
  \begin{minipage}[b]{0.46\textwidth}
    \centering
    \includegraphics[width=\textwidth]{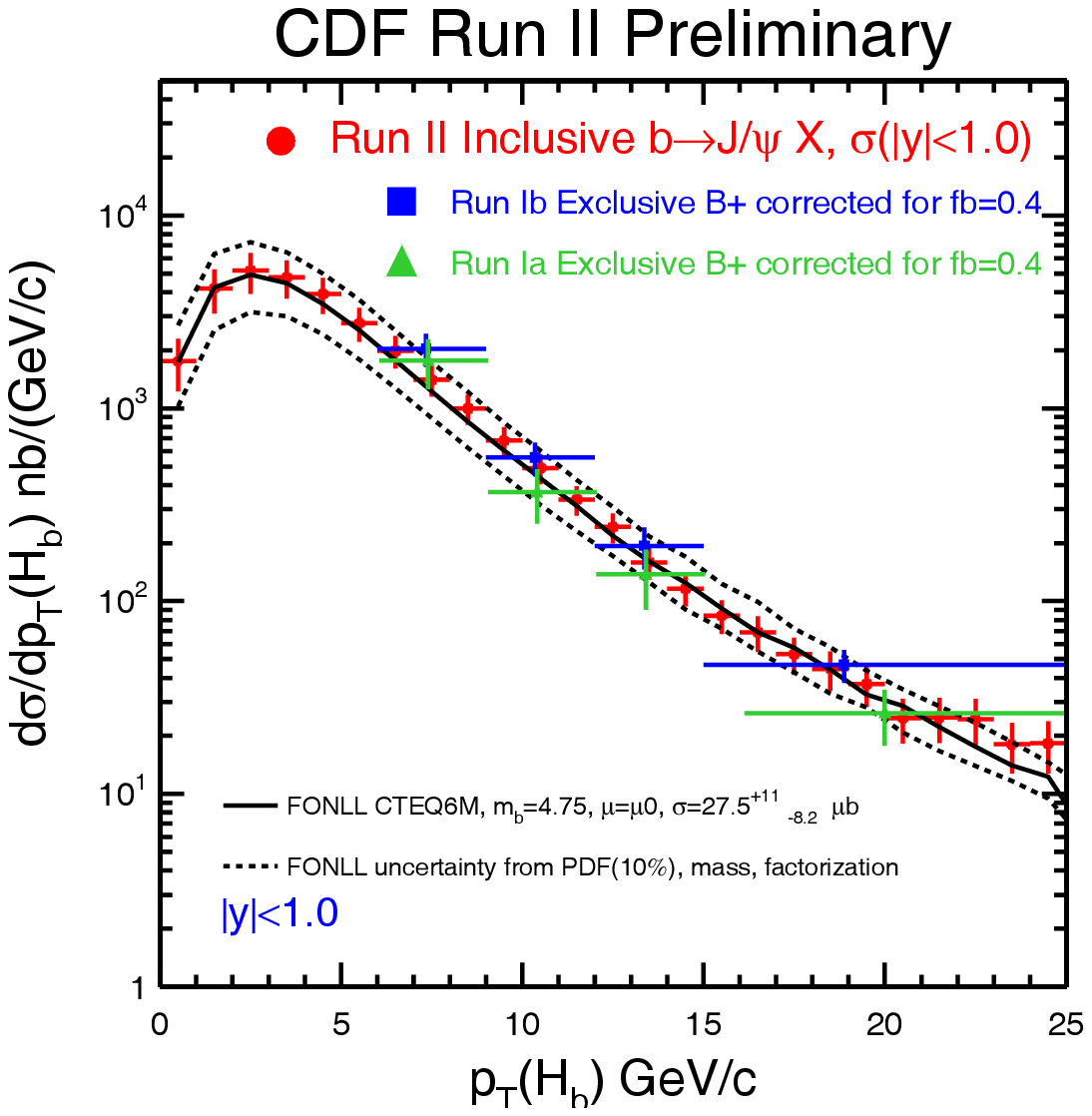}
  \end{minipage}
  \end{center}
  \caption{\it
      The preliminary, CDF \Hadron{H}{B} differential cross-section
      times branching ratio (left)
      and that result, corrected to the \qrk{b}-quark level compared
      with older CDF measurements and the FONLL prediction (right).
    \label{fig:cdfhbxs} }
\end{figure}

These new calculations \cite{cfmnr} have been compared to a
preliminary measurement of the \Hadron{B}{}-hadron cross-section
by CDF, using \Hadron{H}{B}\ra \Jpsi $X$ decays.
CDF selects \Hadron{H}{B} \ra \Jpsi\ decays in the \Jpsi \ra \mupmum\
mode from 37 \unitexp{pb}{-1} of their Run II data 
using the position of
the \Jpsi\ vertex with respect to the primary \qqb{p} interaction point
to distinguish long-lived \Hadron{H}{B} decays from prompt \Jpsi\
production. 
The resulting \Hadron{H}{B} cross-section times branching ratio is
shown on the left side of Figure \ref{fig:cdfhbxs} while a
comparison of this new result to older CDF measurements and to the
FONLL prediction \cite{cfmnr} is shown on the right side. As can be
seen, the agreement between data and prediction is excellent, both in
shape and normalization. The total cross-sections, corrected to the quark
level for the CDF data is
  $\sigma(\qqb{p} \ra \aqrk{b} X, |y_b| < 1.0)$
  = 29.4$\pm$0.6$\pm$6.2 $\mu$b
in remarkable agreement with the FONLL prediction of
  25.0$^{+12.6}_{-8.1}$ $\mu$b.

\begin{figure}[htb]
  \onefig{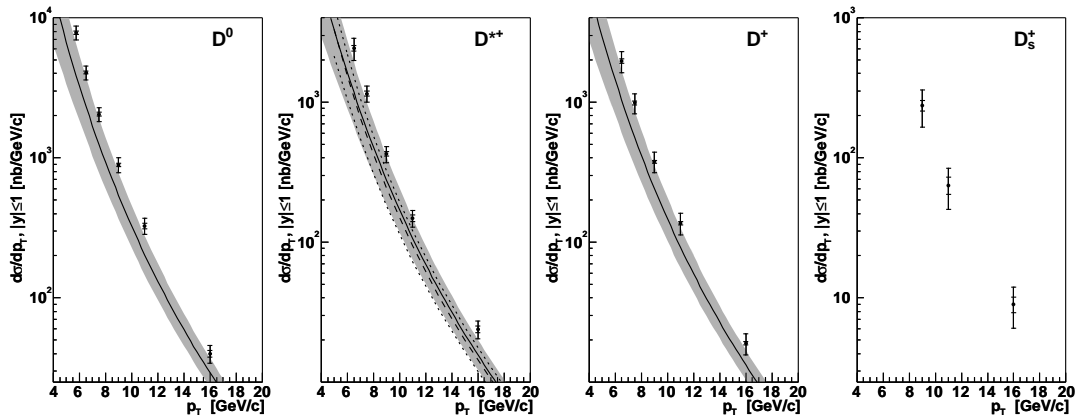}
  \caption{\it
      The CDF differential charm meson cross-section measurements
      compared to the FONLL prediction.
    \label{fig:cdfcxs} }
\end{figure}

New CDF measurements of open charm production \cite{cdfcxs}
have also been compared
to FONLL predictions \cite{caccnasc}.
The measurement uses 5.8 \unitexp{pb}{-1}
of hadronic charm decay triggers
collected with the CDF Silicon Vertex Trigger. 
Prompt contributions to the sample of reconstructed
\Hadronup{D}{0}, \Hadronup{D}{*+}, \Hadronup{D}{+} 
and $D_s^+$ mesons are obtained using the impact
parameter of the charm meson candidate.
The measured differential cross-sections in the rapidity region $|y|
\leq$1 agree fairly well with FONLL predictions,
as shown in Figure \ref{fig:cdfcxs},
although the data lie systematically on the high side of the theory.

\subsection{Open Beauty and Charm at Fixed Target}
The Hera-B experiment has made new measurements of open \qrk{b}- and
\qrk{c}-quark production in a fixed target environment. Their
preliminary measurement of the \qqb{b} cross-section
uses \Jpsi \ra \epem , \mupmum\
decays with displaced vertices from a total of 320K \Jpsi\ candidates
in $\sim$35\% of their 02-03 data sample. 
The new Hera-B
measurement,
$\sigma(\qqb{b})$ = 12.3$^{\mathrm{+3.5}}_{\mathrm{-3.2}}$ 
(stat) nb/nucleon,
is lower than their previous result,
32$^{\mathrm{+14\ +6}}_{\mathrm{-12\ -7}}$ nb/nucleon \cite{herabold}, 
which used the 40-times smaller 2000 data sample.
It agrees well with the prediction of Kidonakis, \etal\
\cite{kidherab}, 30$\pm$13 nb/nucleon,
although the errors on the prediction are still rather large.

Hera-B has also made a preliminary measurement of the open charm
cross-section using 98$\pm$12 \Hadronup{D}{0},
189$\pm$20 \Hadronup{D}{+} and 43$\pm$8 \Hadronup{D}{*+} 
fully reconstructed mesons.
The resulting $D\bar{D}$ cross-sections,
in $\mu$b/nucleon,
extrapolated to the full $x_F$ range, are
$\sigma(D^+)$ = 30.2$\pm$4.5$\pm$5.8 and
$\sigma(D^0)$ = 56.3$\pm$8.5$\pm$9.5,
which are consistent with
previous measurements but significantly more accurate.

Measurements of charm production have also been made by the FOCUS
collaboration, which has produced new results on charm
baryon/anti-baryon production asymmetries \cite{focusasym}.
The asymmetry is predicted to be vanishingly small by
perturbative QCD. However ``leading particle effects'', can enhance
the production of baryons sharing valence quarks with the target or
projectile particles.
The measured integrated and differential asymmetries 
for $\Lambda_c^+$, $\Lambda_c^+$(2625), $\Sigma_c^{++(*)}$ and 
$\Sigma_c^{0(*)}$
agree poorly with predictions from the PYTHIA Monte Carlo.
For example, FOCUS measures the asymmetry in the production of 
$\Lambda_c^+$ baryons and anti-baryons to be
0.111$\pm$0.018$\pm$0.012,
1.8$\sigma$ away from the prediction of 0.073.
A better description of older
asymmetry measurements has been achieved using heavy quark
recombination models \cite{qrecomb}.
But this has yet to be compared to the FOCUS data.

\subsection{Open Beauty and Charm in \ep\ Collisions}
Experimentally, both H1 and ZEUS search for \qrk{b}-quark production
in \ep\ collisions
using muon plus jet(s) events. ZEUS separates \qrk{b}-quark events from
backgrounds using the component of the muon's momentum transverse the
the closest jet axis, \Ptrel , while H1 takes advantage of their
silicon strip detectors to include the impact parameter of the muon
track with respect to the primary interaction vertex, along with
\Ptrel , to their list of discriminating variables.

\begin{figure}[htb]
  \threefigs{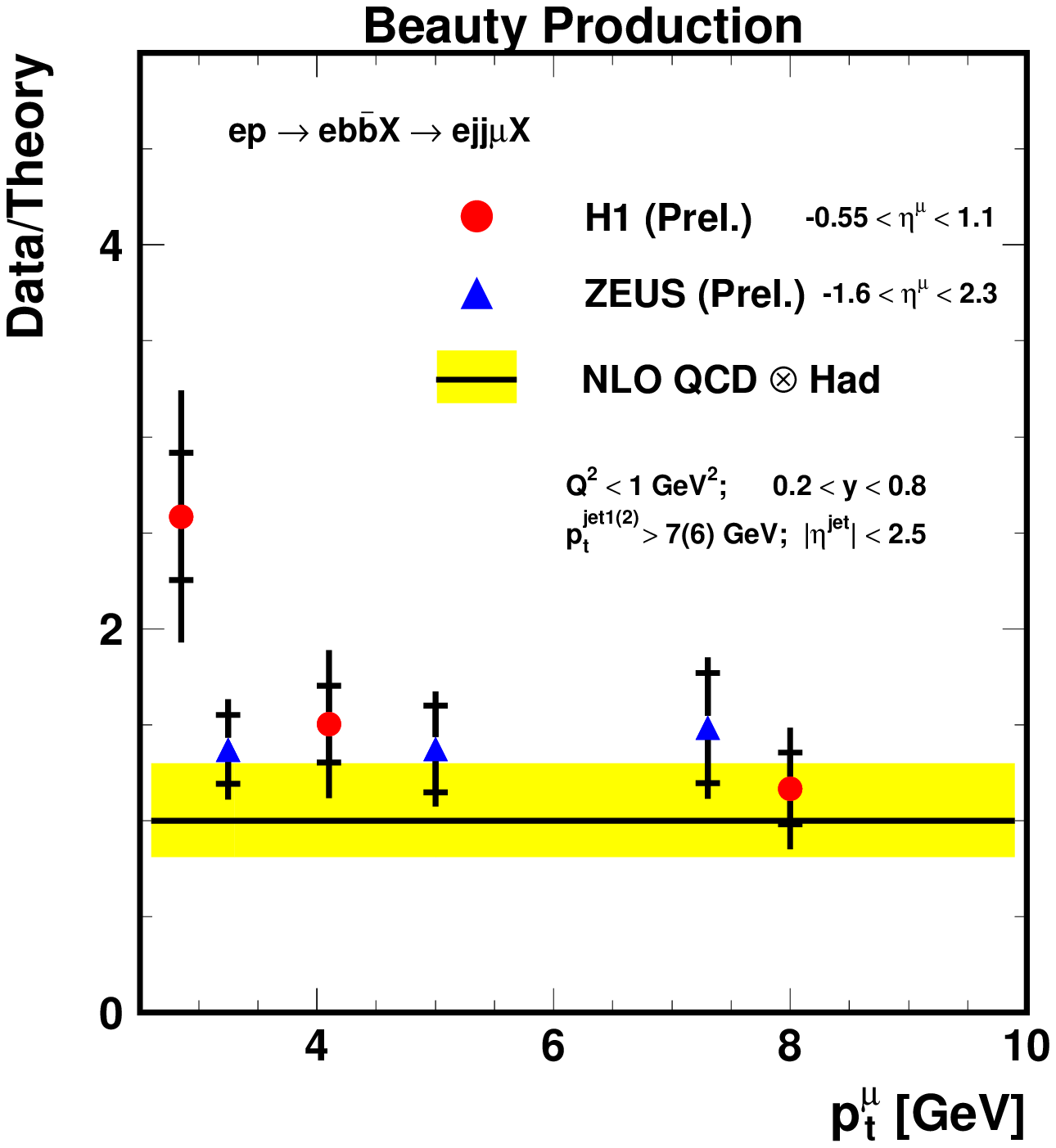}{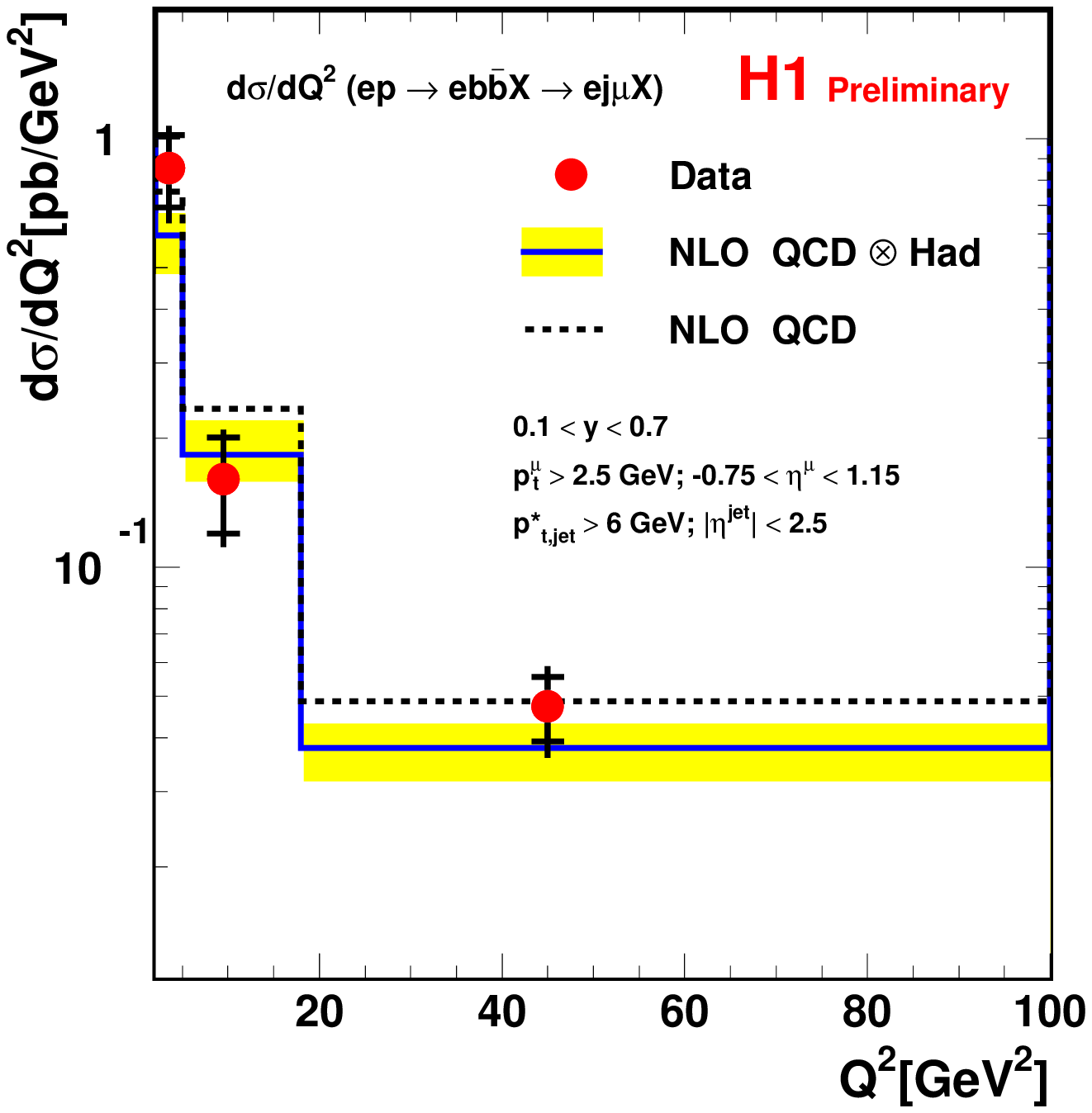}{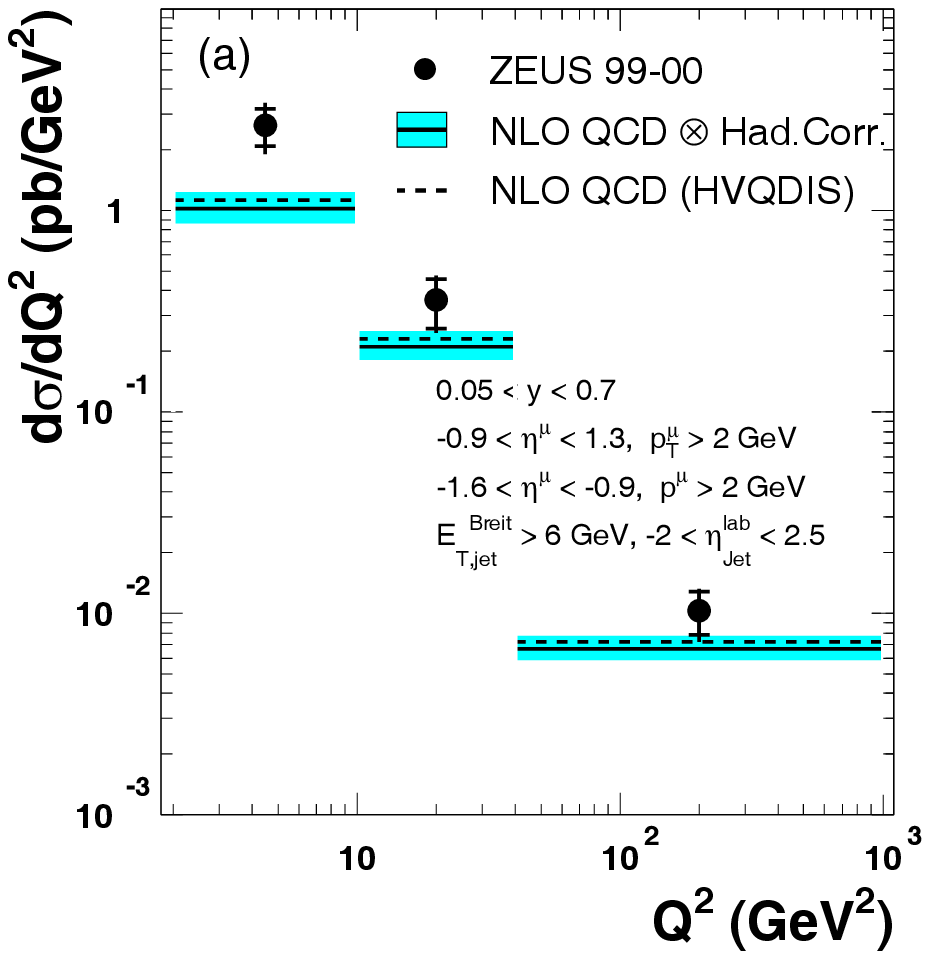}
  \caption{\it
      Measurements of \qrk{b}-production by H1 and ZEUS in
      photo-production (left plot) and DIS
      (middle and right plots).
    \label{fig:herab} }
\end{figure}

The experiments make measurements in two kinematic
regions -- the deep inelastic scattering (DIS) regime, where photon
virtuality is high (\Qsq\ $>$ 1 \unitexp{GeV}{2})
and the photo-production (PhP) regime, where there the photon is
nearly real (\Qsq\ $<$ 1 \unitexp{GeV}{2}).
Each of these regimes is sensitive to different effects in heavy quark
production and provide complementary input to the measurements from
the Tevatron, where log(\pT /$m$) effects, for example,
are expected to be much more important. 
The results of preliminary H1 measurements from 2003 and 2004 and of
published ZEUS data \cite{zeusbxs} are shown in
Figure \ref{fig:herab}.

\begin{figure}[htb]
  \twofigs{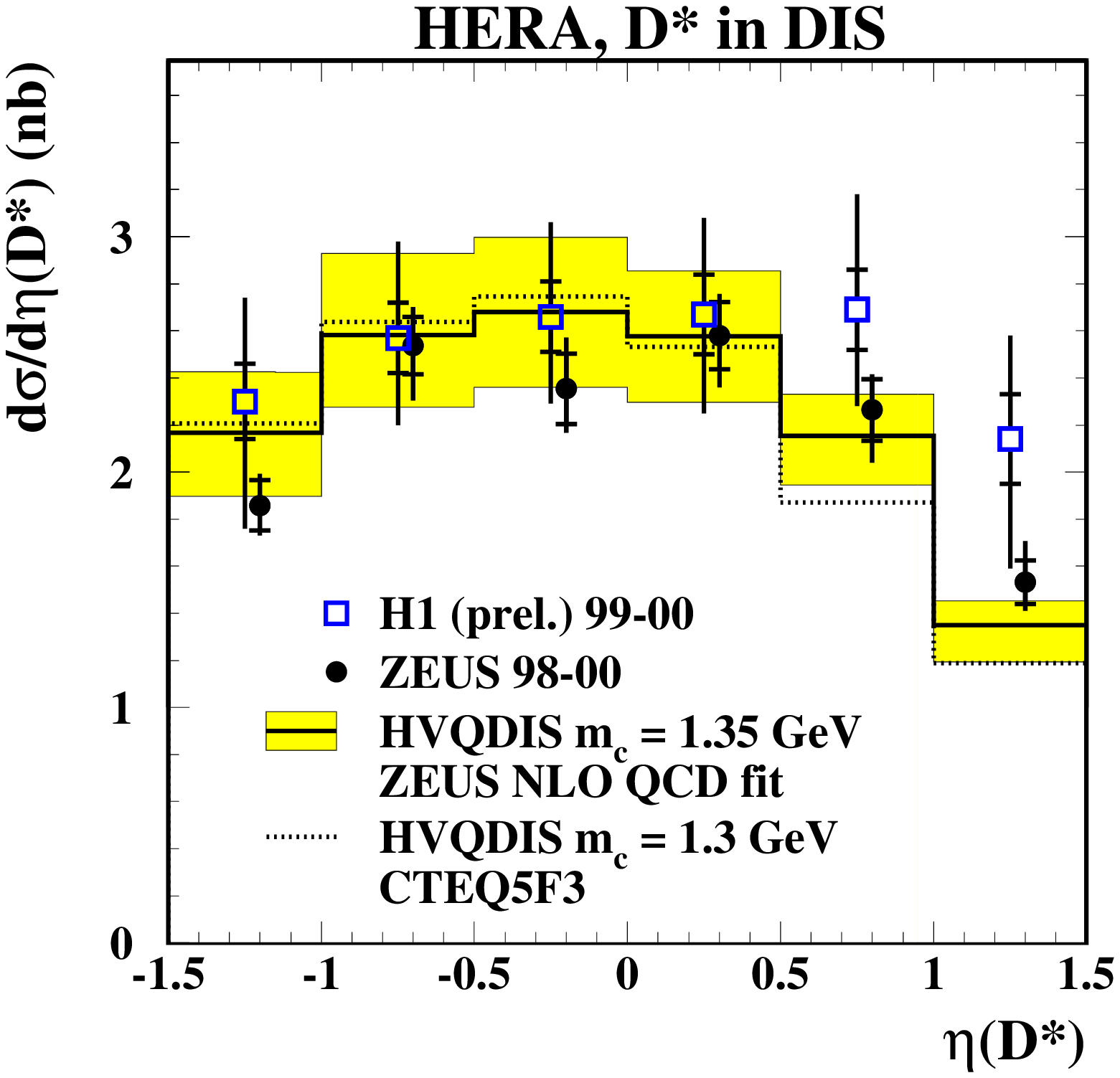}{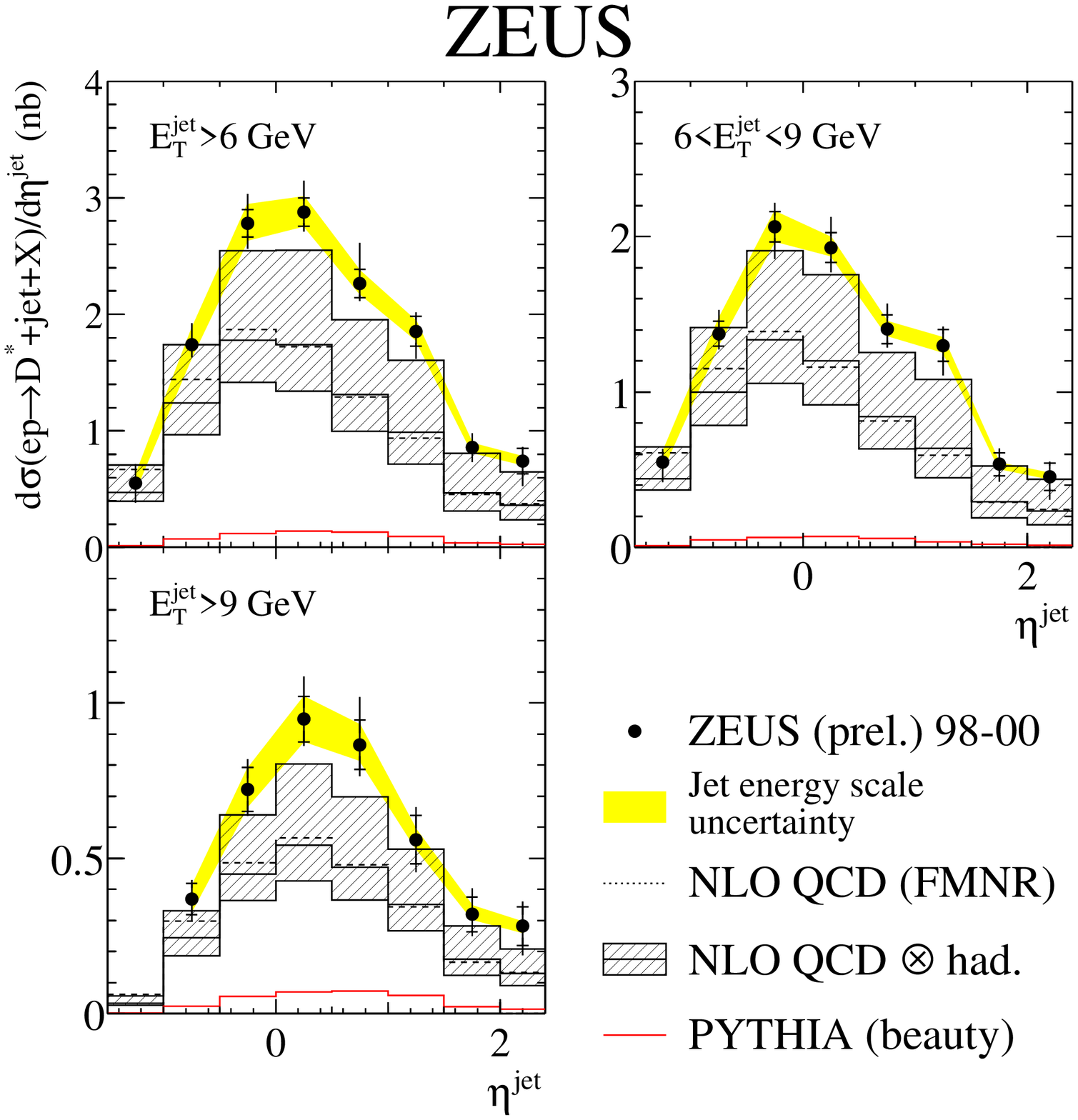}
  \caption{\it
      Measurements of \Hadronup{D}{*} production by H1 and ZEUS in
      DIS (left) and by ZEUS in photo-production (right)
    \label{fig:herac} }
\end{figure}

Charm quark production is also measured by both experiments using
\Hadronup{D}{*\pm } mesons. 
Preliminary results from H1 (2003) and from ZEUS (2002) in PhP events
as well as recently published ZEUS data \cite{zeusdstdis}
are presented in Figure \ref{fig:herac}. 

Agreement between both beauty and charm data and NLO predictions is
generally good within the relatively large experimental and
theoretical errors. Some problems may arise in \qrk{b}-quark
production at low \pT\ and low \Qsq\
(see Figure \ref{fig:herab}).
However, H1 and ZEUS do not see the same discrepancies.
In the charm PhP data, mild deviations between data and theory are
observed in the medium \pT\ and high $\eta$ regions.
ZEUS has studied these further in a preliminary measurement of the
cross-section of jets containing \Hadronup{D}{*\pm} mesons, designed
to reduce sensitivity to hadronization effects. As can be seen in
Figure \ref{fig:herac} some disagreement between data and predictions
remains indicating that hadronization is unlikely to be the 
main cause of the problem.

\subsection{\qrk{b}- and \qrk{c}-Quark Summary}
The general picture emerging from new measurements of beauty and charm
production and from recent theoretical advances is of remarkably
better agreement between data and theory.
Comparisons between measurement and prediction for the
results discussed above show agreement to within about two sigma
(taking into account both experimental and theoretical errors) for all
recent measurements.
This is obviously a big improvement over the situation a few years ago.
However, optimism should not be allowed
to run rampant over caution. 
Uncertainties on nearly all measurements are
dominated by systematic errors indicating that higher statistics alone
will be unlikely to produce major improvements in accuracy. On the
theoretical side, uncertainties on the predictions are nearly always
substantially larger than those on the measurements further adding to
the difficulty of making quantitative comparisons.

\section{Heavy Quarkonium Production}
\subsection{More History}
As was the case with open beauty and charm, our
understanding of the production of bound heavy quark-antiquark states has
had a checkered past. 
(For a discussion of the decays of quarkonia see \cite{galik}).
Until the late 90's the direct production of \Jpsi\
and \Ups\ states was expected to proceed via a color singlet mechanism
(CSM) where the \qqb{Q} meson retains the quantum numbers of the
\qqb{Q} pair produced in the hard scatter. CDF measurements of prompt
\Jpsi\ and \psipr\ production in Run I \cite{cdfpsixs} were higher than
CSM predictions by a factor of 50 though.
This discrepancy was largely resolved by the introduction of the color
octet model (COM) of quarkonium production \cite{com}. This model allows
contributions from the
production of \qqb{Q} pairs in a color octet state, which evolve into
color singlet states by the emission of a soft gluon.
The COM also improved the agreement between the rate of \Jpsi\
production observed in \ep\ collisions and predictions 
\cite{h1psiphp,h1psidis,zeuspsiphp}.

However, the introduction of the COM has a price:
unlike the CSM, the COM predicts large values for the polarization of
quarkonia states at high \pT . These large polarizations have not been
observed experimentally in \Jpsi\ or \psipr\ production at the
Tevatron \cite{cdfpsipol} or fixed target experiments.
Measurements of \Ups\
polarization tend to suffer from limited statistics and are generally
not yet 
able to discriminate significantly between CSM and COM predictions for
polarization. 
However, the NuSea collaboration finds large
polarization for \Ups (2S,3S) states \cite{nuseaups}, in agreement
with COM predictions.

\subsection{\Jpsi\ Polarization at Fixed Target}
The NuSea collaboration has recently turned to the \Jpsi ,
with a new polarization measurement of those mesons 
in proton--copper collisions \cite{nusea}.
Approximately nine million \Jpsi \ra
\mupmum\ candidates are selected allowing
measurements of the polarization to be made in several bins of
$x_F$.
An average polarization of
0.069$\pm$0.004$\pm$0.080 is found, which agrees
with previous fixed target findings of very small polarization,
but with substantially better accuracy.
The measurement is lower than
predictions based on the COM, which range from 0.35 to 0.65.
But \Jpsi\ mesons produced in decays of other particles (predicted
to have small polarizations) have not been excluded from this analysis,
or from most of the other fixed target results. So direct
comparisons with COM predictions are difficult.

\subsection{\Jpsi\ Production in \ep\ Collisions}
\begin{figure}[htb]
  \begin{center}
  \begin{minipage}[b]{0.52\textwidth}
    \centering
    \includegraphics[width=\textwidth]{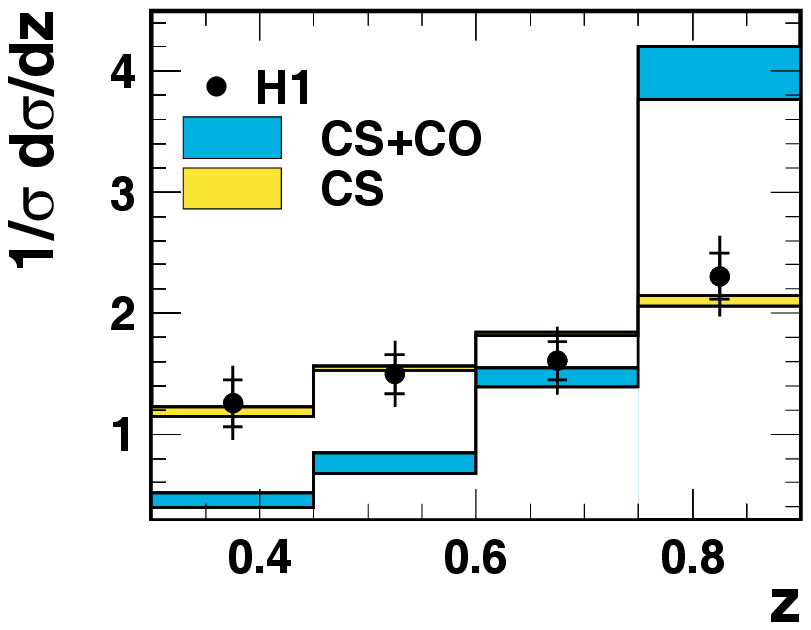}
  \end{minipage}
  \begin{minipage}[b]{0.38\textwidth}
    \centering
    \includegraphics[width=\textwidth]{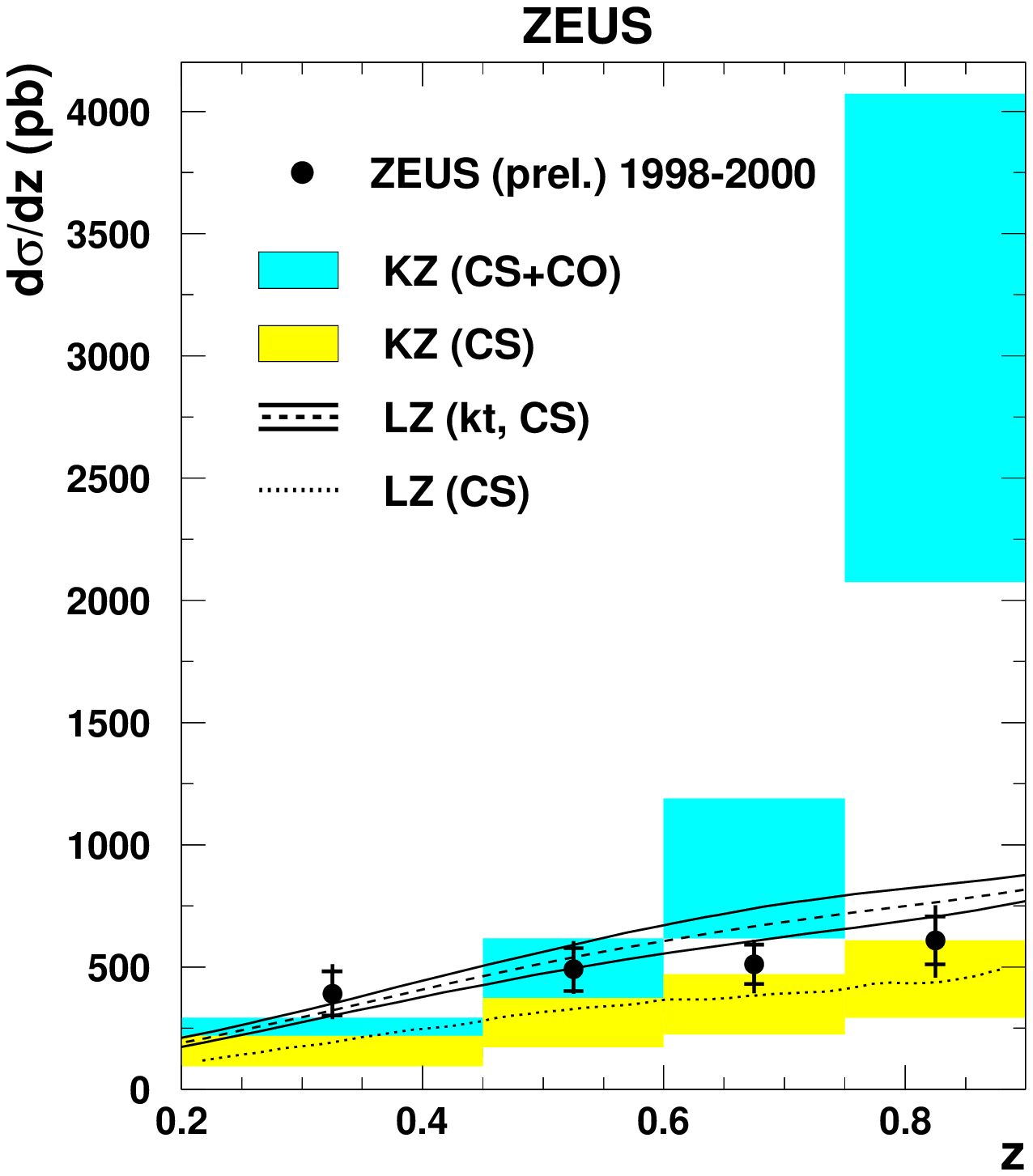}
  \end{minipage}
  \end{center}
  \caption{\it
      A comparison of H1 and ZEUS \Jpsi\ differential cross-sections
      in DIS with CSM and COM predictions.
    \label{fig:herapsi} }
\end{figure}

The ZEUS collaboration has released recent, preliminary results on the
production of \Jpsi\ mesons in DIS events and their polarization in a
PhP sample. While the polarization measurement has too low statistics
to allow a distinction to be made between CSM and COM predictions, the
DIS production measurement does have sensitivity to differences
between the models.
This measurement selects 203$\pm$19 \Jpsi \ra \mupmum\ decays out
of 73 \unitexp{pb}{-1} of data and can be
compared to a previously published H1 result \cite{h1psidis} where
458$\pm$30 \Jpsi \ra \mupmum , \epem\ decays were observed in 77
\unitexp{pb}{-1} of data. Measurements of the differential
cross-section are shown in Figure \ref{fig:herapsi} for both the ZEUS
and H1 data. These data imply that the shape of the cross-section is
better modeled by the CSM than by the COM, although errors on the
predictions are still quite large.

\subsection{Heavy Quarkonium Summary}
Despite recent measurements, our view of the production of heavy
quarkonia states remains obscured.
A COM description of the data is strongly preferred by measurements of
\Jpsi\ production at the Tevatron and, to a lesser extent by \Ups\
polarization measurements at fixed target.
On the other hand, the color singlet model provides a better
description of \Jpsi\ polarization at the Tevatron and fixed target
experiments as well as matching the shape of the \Jpsi\ differential
production cross-section in DIS events at HERA.
Finally, the absolute normalization of the \Jpsi\ cross-section in DIS
and PhP events is described well be neither model.
As is the case with open beauty and charm production though,
quarkonium measurements tend to be systematics limited and
uncertainties on theoretical predictions are quite large.

\section{New Particles}
\subsection{The \Xpart\ at the Tevatron}
In the summer of 2003, the Belle collaboration announced the
observation of a new particle with a mass of around 3872 MeV
in \Hadronup{B}{+} \ra\ \Hadronup{K}{+}\Xpart\ decays \cite{bellex}.
This particle,
which like the \psipr ,
decays to \Jpsi \pip \pim\
has now been observed by several other experiments \cite{galik}
including CDF \cite{cdfx} and \Dzero\ \cite{d0x}.
Both Tevatron experiments
observe large signals, with CDF finding 730$\pm$90
(11.6$\sigma$) events with a fitted mass of
3871.3$\pm$0.7$\pm$0.3 MeV
and \Dzero\ seeing 522$\pm$100 (5.2$\sigma$) with a fitted
mass of 
3871.8$\pm$3.1$\pm$3.0 MeV (referenced to the \psipr\ mass).
The large signal samples available to the Tevatron experiments
(the original observation by Belle consisted of $\sim$35 signal events)
will allow detailed studies of the \Xpart\ to be made.
\Dzero\ has started this process by studying 
several kinematic properties, in production and decay, of their \Xpart\
sample, finding that the \Xpart\ behaves very much like the \psipr\
within the statistics of their test.

\subsection{Charmed Pentaquarks?}
\begin{figure}[htb]
  \twofigs{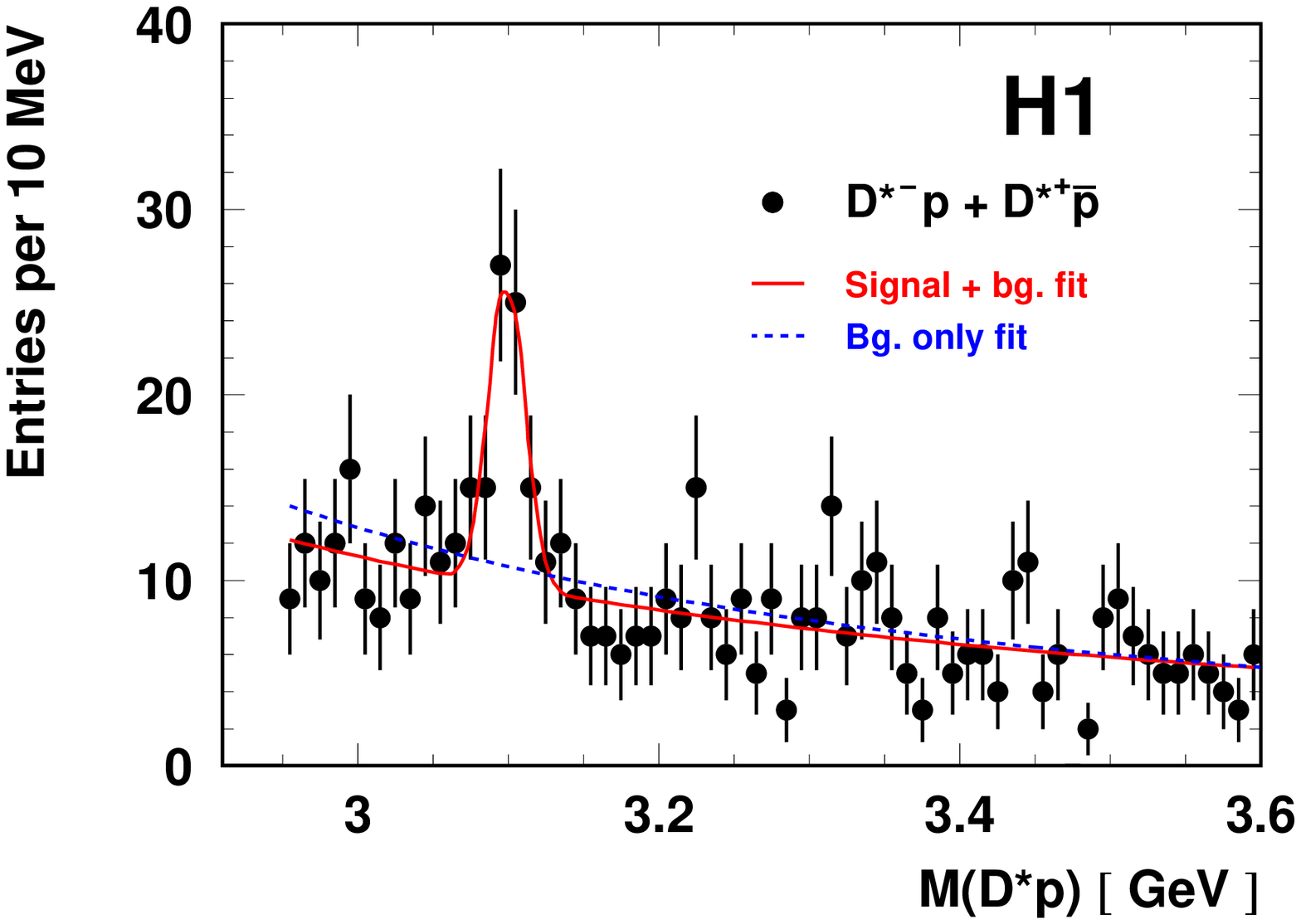}{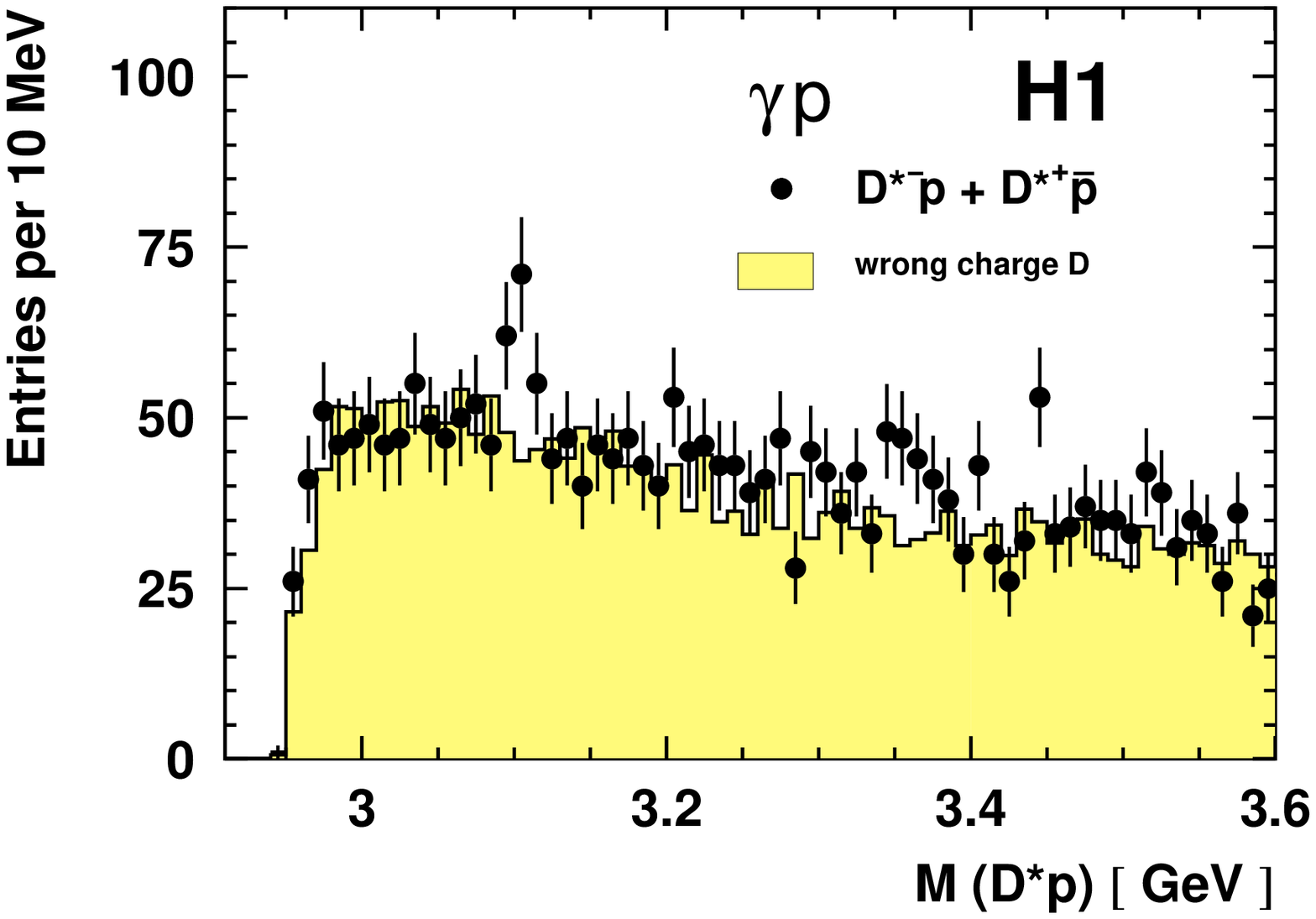}
  \caption{\it
      Evidence for the \cpentaq\ by
      H1 in DIS (left) and photo-production (right).
    \label{fig:thetac} }
\end{figure}

Controversy continues to boil over the evidence for a pentaquark particle,
\pentaq , with a valence quark content of ($uudd\bar{s}$)
\cite{nakano}. 
Undeterred by this uncertainty, several groups have looked for a
charmed pentaquark, \cpentaq , with quark content ($uudd\bar{c}$).
The H1 collaboration sees evidence for this particle in the decay
\cpentaq \ra\ \Hadronup{D}{*-}\qrk{p} \cite{h1pentaq}.
As shown in Figure \ref{fig:thetac}, significant signals are seen by
H1 in both DIS and PhP. They find 51$\pm$11 (5.4$\sigma$) \cpentaq\
candidates at a mass of 3099$\pm$3$\pm$5 MeV
from a sample of $\sim$8500 \Hadronup{D}{*} mesons
in 75 \unitexp{pb}{-1} of data.

The primary experimental difficulty in the H1 analysis is to avoid
reflections from \Hadronup{D}{**}\ra \Hadronup{D}{*}$\pi$ decays,
which peak in the 3100 MeV region if the pion is misidentified as a
proton. H1 avoids these reflections by separating pions from protons
using dE/dx. They have performed many cross-checks to verify
the reliability of this selection.

Motivated by H1's result, ZEUS, CDF and FOCUS have conducted
preliminary searches for the \cpentaq . Despite having 
similar sensitivity to H1 and
larger samples
of \Hadronup{D}{*} mesons -- 43K, 200K and 36K for ZEUS, CDF and FOCUS,
respectively --
none of these experiments observe any evidence for the H1 signal.

\subsection{New Particles at SELEX}
\begin{figure}[htb]
  \threefigs{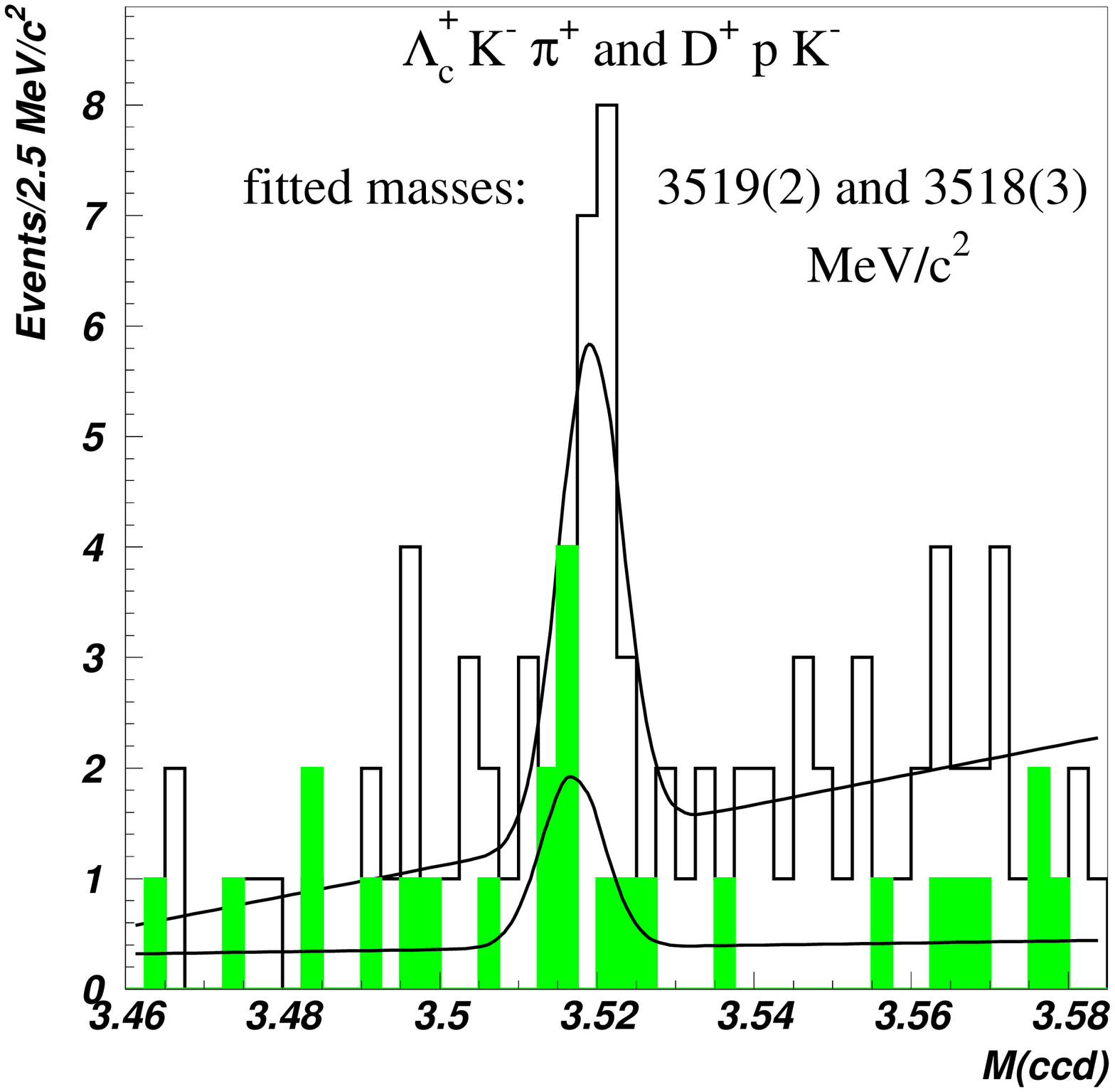}{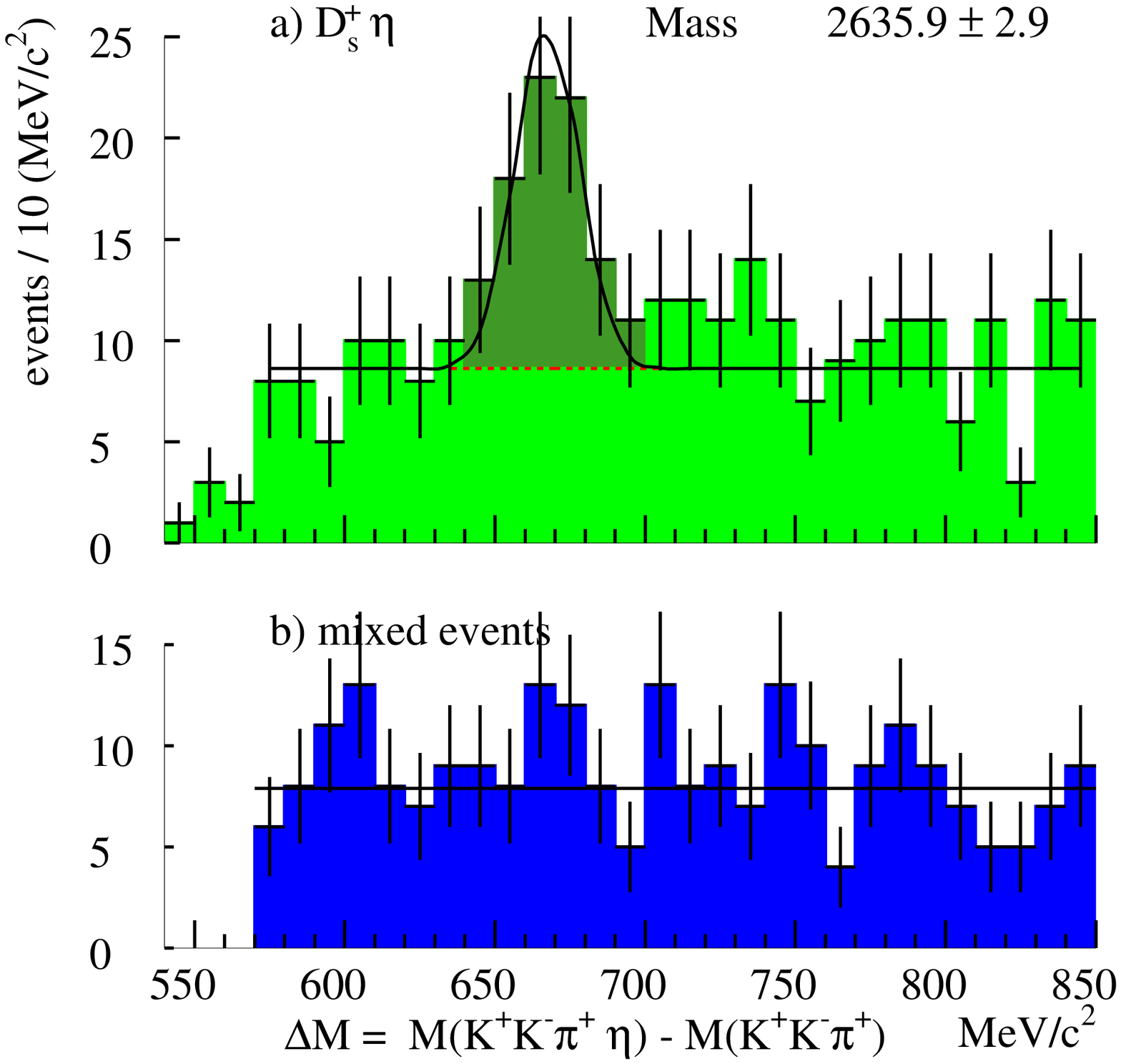}{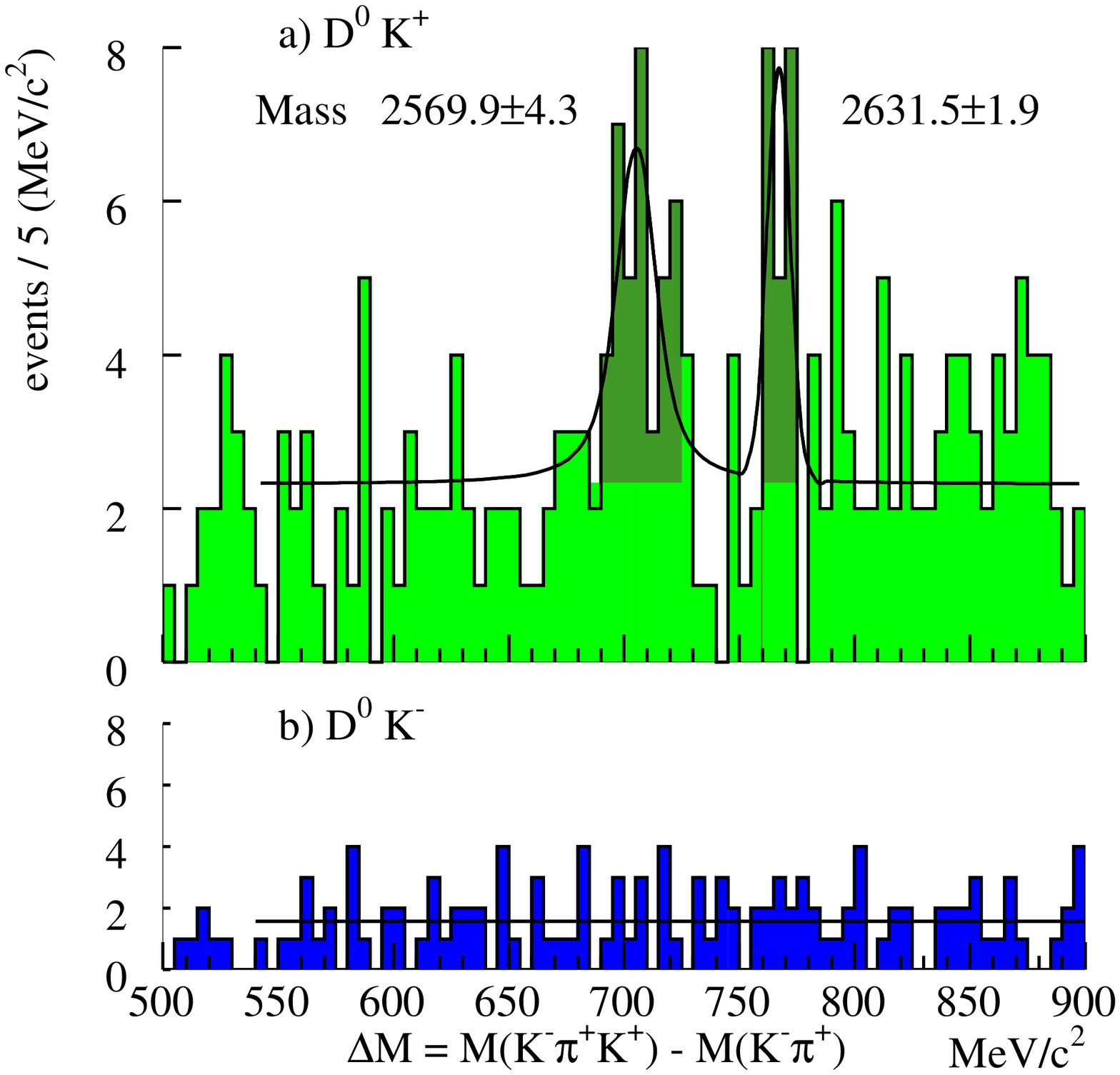}
  \caption{\it
      SELEX mass plots for the 
      \xicc\ (left),
      the \DsJ \ra $D_s^+ \eta$ (middle)
      and the \DsJ \ra $D^0 K^+$ (left).
    \label{fig:selex} }
\end{figure}

Two new particles have recently been sighted by the SELEX
collaboration. Significant mass peaks for a doubly charmed baryon,
\xicc , decaying to $\Lambda_c^+ K^- \pi^+$ and $p D^+ K^-$ \cite{selexxicc}
and a charm-strange meson, \DsJ ,
decaying to $D_s^+ \eta$ and $D^0 K^+$ \cite{selexdsj} 
are shown in Figure \ref{fig:selex}.
The new measurement of the \xicc\ supports a previous SELEX
observation of this particle, but has not been confirmed by the FOCUS
or E791 collaborations. 

Should the evidence for these particles hold up to further scrutiny,
they promise to provide some interesting physics. Both have rather
strange properties. The decay length distribution of the \xicc\
candidates indicates a lifetime significantly shorter than expected
and the relative branching ratios of the two observed decay modes are
inconsistent with phase space expectations. The two \DsJ\ decay modes
observed also show a large difference from phase space predictions
and, even more mysteriously, the width of the particle is much
narrower than expected, $<$17 MeV at 90\% C.L.

\section{Where to Now?}
After several years of particularly intense activity in the area of
heavy quark production, our understanding of the topic has increased
substantially. Problems that have plagued the comparison of
\qrk{b}-quark data and predictions seem to have been largely resolved
thanks to the efforts of both experimentalists and theorists.

The field should, by no means, slide into complacency though. Both
experimental systematic errors and theoretical uncertainties in beauty
and charm production must be reduced before modeling of these
processes can approach the level needed for understanding the next
round of results from the LHC. Confusion also continues to reign in
the area of heavy quarkonium production. Seemingly inconsistent
experimental results across production and polarization measurements
need to be resolved.
And, as we have seen, surprising new particles, possibly pointing the
way to interesting new phenomena, are waiting in the wings.

Fortunately, the future of heavy flavor production physics looks
bright. Both the Tevatron and HERA accelerators have started new runs,
which promise orders of magnitude more data than currently available,
with upgraded detectors. Further down the road the LHC experiments,
Atlas, CMS and LHCb, as well as BTeV at Fermilab should be able to
collect heavy flavor data sets that dwarf those foreseen from Run II
at the Tevatron and HERA, allowing production studies using exclusive
final states. The optimism engendered by this possibility
must be tempered by the knowledge that the physics goals of the
upcoming experiments are not aimed primarily at heavy flavor
production. In particular, the ability to do this type of physics will
be limited by the performance of trigger systems, which rely
primarily on muon triggers to collect heavy flavor data. BTeV, with
its displaced vertex trigger, is an exception, which deserves special
attention here.
Despite the challenges a heavy flavor production program presents,
though, 
active efforts have been started to study its possibilities
\cite{heralhc}. I believe that we can look forward to exciting reviews
of heavy flavor production for many {\em Physics in Collision}
conferences to come.

\section{Acknowledgments}
The material presented here is the result of the sweat and toil of an army
of physicists. Shamefully, I cannot acknowledge all who participated
by name. But I benefited particularly from the advice of H. Cheung,
M. Corradi, R. Galik, V. Jain, R. Jesik, G. Landsberg, M. Leitch,
M. zur Nedden, P. Newman, C. Paus, J. Russ, L. Silvestris,
M. Smizanska, K. Stenson, A. Zieminski, D. Zieminska and
A. Zoccoli. I would also like to thank the conference organizers for
giving me an excuse to come to Boston and for putting together such a
stimulating three days.


\end{document}